\documentclass[journal,draftclsnofoot,onecolumn,12pt]{IEEEtran}

\usepackage{acronym}
\usepackage{amsfonts}
\usepackage[dvips]{graphicx}
\usepackage{times}
\usepackage{cite}
\usepackage{amsmath}
\usepackage{array}
\usepackage{amssymb}
\usepackage{stfloats}
\usepackage{diagbox}
\usepackage{graphicx}
\usepackage{footnote}
\usepackage{amsthm}
\usepackage{booktabs}
\usepackage{array}
\usepackage[ruled,vlined]{algorithm2e}
\usepackage{subeqnarray}
\usepackage{cases}
\usepackage{threeparttable}
\usepackage{color}
\usepackage{epstopdf}
\usepackage{multirow}
\usepackage{tabularx}
\usepackage{enumerate}
\usepackage{multicol}
\usepackage{hyperref}
\usepackage{subfig}
\usepackage{caption}

\def\bb0{{\mathbb{0}}}

\def\bb{{\mathbf{b}}}

\def\b0{{\mathbf{0}}}

\def\sf0{{\mathsf{0}}}

\def\rm0{{\mathrm{0}}}

\acrodef{CSI}[CSI]{channel state information}
\acrodef{CSIT}[CSIT]{channel state information at the transmitter}
\acrodef{CSIR}[CSIR]{channel state information at the receiver}
\acrodef{MIMO}[MIMO]{multiple-input multiple-output}
\acrodef{SISO}[SISO]{single-input single-output}
\acrodef{MISO}[MISO]{multiple-input single-output}
\acrodef{SIMO}[SIMO]{single-input multiple-output}
\acrodef{ADCs}[ADCs]{analog-to-digital convertors}
\acrodef{SNR}[SNR]{signal-to-noise ratio}
\acrodef{AWGN}[AWGN]{additive white Gaussian noise}
\acrodef{MRT}[MRT]{maximal ratio transmission}
\acrodef{DFT}[DFT]{Discrete Fourier Transform}
\acrodef{ULA}[ULA]{uniform linear array}
\acrodef{UPA}[UPA]{uniform planar array}
\acrodef{LS}[LS]{least squares}
\acrodef{ALMMSE}[ALMMSE]{approximate linear minimum mean squared error}
\acrodef{QIHT}[QIHT]{quantized iterative hard thresholding}
\acrodef{QIST}[QIST]{quantized iterative soft thresholding}
\acrodef{SVD}[SVD]{singular value decomposition}

\usepackage{float}
\ifCLASSINFOpdf

\else
\fi
\begin{document}

\title{Wireless Communications with Programmable Metasurface: New Paradigms, Opportunities, and Challenges on Transceiver Design}

\author{Wankai Tang, Ming Zheng Chen, Jun Yan Dai, Yong Zeng, \\
Xinsheng Zhao, Shi Jin, Qiang Cheng and Tie Jun Cui

}

\maketitle
\vspace{-1cm}
\begin{abstract}

Many emerging technologies, such as ultra-massive multiple-input multiple-output (UM-MIMO), terahertz (THz) communications are under active discussion as promising technologies to support the extremely high access rate and superior network capacity in the future sixth-generation (6G) mobile communication systems. However, such technologies are still facing many challenges for practical implementation. In particular, UM-MIMO and THz communication require extremely large number of radio frequency (RF) chains, and hence suffering from prohibitive hardware cost and complexity. In this article, we introduce a new paradigm to address the above issues, namely wireless communication enabled by programmable metasurfaces, by exploiting the powerful capability of metasurfaces in manipulating electromagnetic waves. We will first introduce the basic concept of programmable metasurfaces, followed by the promising paradigm shift in future wireless communication systems enabled by programmable metasurfaces. In particular, we propose two prospective paradigms of applying programmable metasurfaces in wireless transceivers: namely RF chain-free transmitter and space-down-conversion receiver, which both have great potential to simplify the architecture and reduce the hardware cost of future wireless transceivers. Furthermore, we present the design architectures, preliminary experimental results and main advantages of these new paradigms and discuss their potential opportunities and challenges toward ultra-massive 6G communications with low hardware complexity, low cost, and high energy efficiency.

\end{abstract}


%
\IEEEpeerreviewmaketitle

\section{Introduction}
2019 is regarded as the first year of the fifth-generation (5G) mobile communication era and it is anticipated that 5G commercial service will be launched in worldwide scale from 2020. Massive multiple-input multiple-output (MIMO) technology has already become a reality, where massive MIMO base stations (BSs) with full-digital transceivers are being commercially deployed and tested in several countries. Looking forward to the future wireless communication technologies for beyond 5G, one possibility is to further scale up in terms of the number of antennas and/or frequency band to meet the ever-increasing traffic demands. Several research institutes and companies have already kicked off the preliminary research attempt on the sixth-generation (6G) mobile communication \cite{6G}. Among the various new technologies that are being actively discussed for the physical layer of 6G mobile systems, the introduction of extremely large aperture antenna arrays (e.g. extremely large aperture massive MIMO (xMaMIMO) \cite{ExtremelyLarge}, ultra-massive MIMO (UM-MIMO) \cite{UMMIMO} and large intelligent surface (LIS) \cite{LIS}), and the use of the terahertz (THz) band \cite{UMMIMO} are regarded as the most promising ones. However, such technologies, though theoretically attractive, are still facing many new challenges that hinder their prospect of practical implementation and future deployment. In particular, due to the extremely large number of radio frequency (RF) chains required by UM-MIMO and the high complexity in the design and manufacture of the high-performance RF components working in high frequency bands, UM-MIMO and THz technologies suffer from the extremely high hardware cost for practical implementation. Note that similar issues also exist in the massive MIMO and millimeter wave schemes of 5G, which will become even more severe in 6G. Similarly, the recently proposed LIS has demonstrated its great potential to enormously increase the system capacity theoretically \cite{LIS}, but there is hardly any literature to discuss the specific hardware implementation to support the signal transmission/reception over such a large continuous aperture. The high hardware costs due to the extremely large number RF chains and/or high working frequency band, together with the high energy consumption and severe heat dissipation issue seriously impede the wide-scale usage of the aforementioned technologies. On the other hand, although the conventional quadrature sampling transceivers with heterodyne or homodyne architecture have been applied in mobile communication systems for many years with great success, these traditional architectures also face many challenges, such as the integration of ultra-massive RF chains and the implementation of THz-band high-performance components, in the future wireless communication systems with UM-MIMO and THz technologies. Therefore, it is of paramount importance to develop flexible and efficient new transceiver architecture to bring the aforementioned technologies into reality. The current digitalization trend of wireless communication systems are evolving from baseband to RF chains and antennas. Researchers in both industry and academia are striving to explore new directions for innovative transceiver technologies, hopefully to redefine the future wireless communication technologies with remarkable breakthroughs.

Against this background, the programmable metasurface is envisioned as a promising technology to address the above-mentioned issues faced by transceiver design in 6G. Programmable metasurface is a new type of artificial electromagnetic (EM) surface, which can be controlled by external signals to realize real-time manipulation of EM waves \cite{MetaNature}. Depending on their actual designs, the programmable metasurfaces are able to control various parameters of EM waves, such as phase, amplitude, and frequency. In this article, by exploiting the superior capability of programmable metasurfaces on manipulating EM waves in a passive manner with low hardware complexity, we propose two promising paradigms for transceiver hardware architectures, namely metasurface-based RF chain-free transmitter and space-down-conversion receiver. Both design paradigms have great potential to reduce the hardware cost and complexity in the future UM-MIMO and THz band communication systems. The rest of the article is organized as follows. Section \ref{Fundamental} provides a brief introduction of programmable metasurfaces. Section \ref{Metatransceiver} presents the architectures and advantages of the proposed metasurface-based RF chain-free transmitter and space-down-conversion receiver, as well as the integrated design of the two. Section \ref{Testsetup} shows our prototype setup and presents the preliminary experimental results. Furthermore, based on these new paradigms and results, we discuss the main challenges and promising future research directions in section \ref{Challenges}. Finally, we conclude the article in section \ref{Conclusion}.

\section{Fundamentals of programmable metasurface}\label{Fundamental}
\begin{figure}[H]
	\centering
	\includegraphics[height=2.51in]{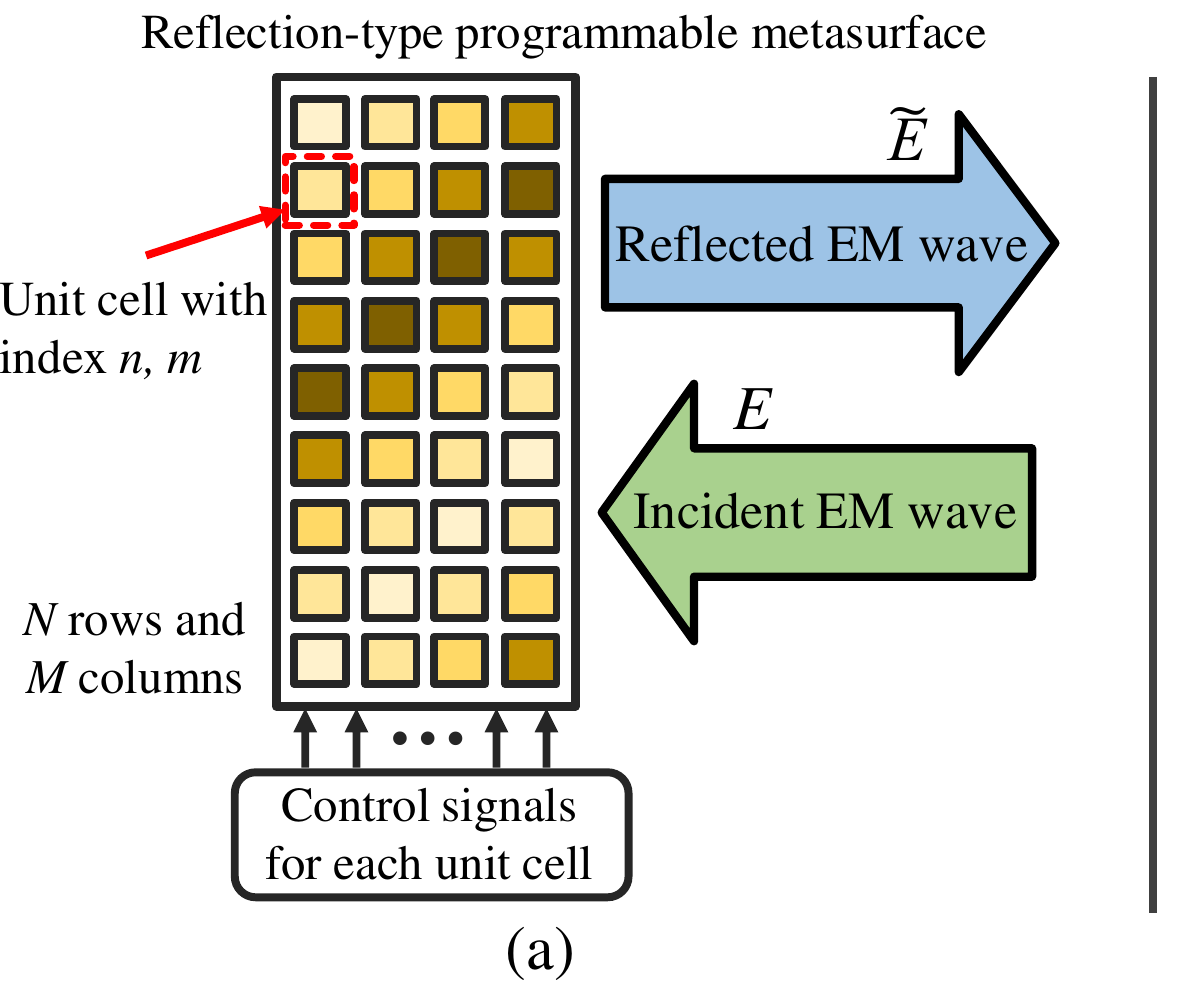}
    \hspace{0.0in}
    \includegraphics[height=2.51in]{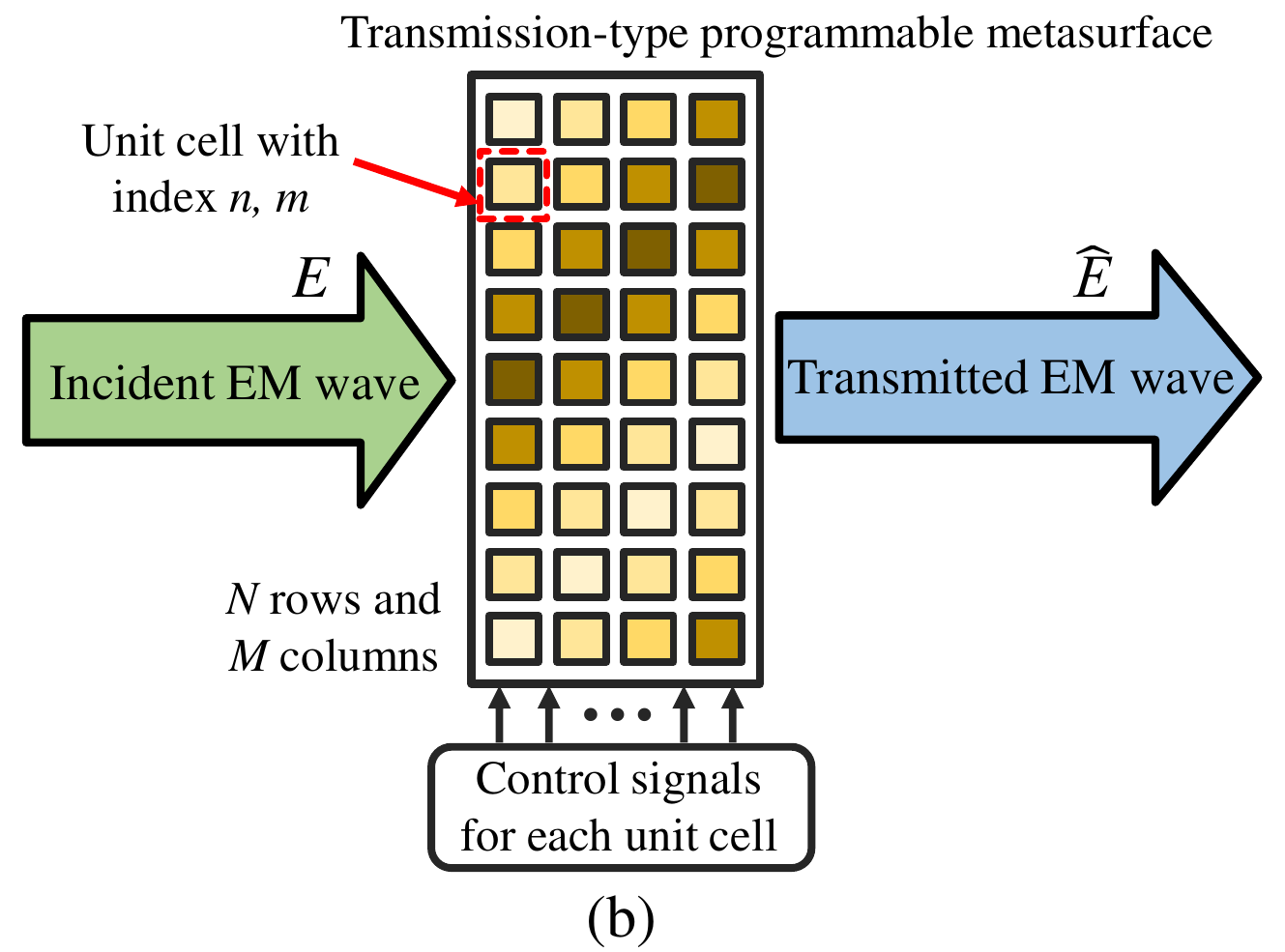}

	\caption{. Schematics of two types of programmable metasurface: (a) reflection type; (b) transmission type.}
	\label{metatype}
\end{figure}

As shown in Fig. \ref{metatype}, programmable metasurfaces essentially consist of two-dimensional artificial sub-wavelength thin-layer structures with programmable electromagnetic properties, which can be potentially applied in a wide range of working frequencies, from microwave to visible light \cite{MetaNature}. They are typically comprised of elaborately designed unit cells with metallic, dielectric and tunable components arranged in a regular array. The electromagnetic parameters of the incident wave, such as the phase and amplitude, can be altered in a programmable manner by controlling the tunable components in each unit cell during the light-matter interaction. This thus provides an interface between the physical electromagnetic world of metasurface and the digital world of information science \cite{MetaInfo}, which is especially appealing for wireless communication applications.

In general, there are two types of programmable metasurfaces, namely reflection-type and transmission-type, as shown in Fig. \ref{metatype}. For reflection-type programmable metasurface, the incident EM wave is converted into the reflected EM wave, with its amplitude and phase adjusted by the external control signals. On the contrary, for transmission-type programmable metasurface, the incident EM wave is mainly converted into the transmitted wave.

Let $N$ and $M$ respectively denote the number of rows and columns of the regularly arranged unit cells in a programmable metasurface. Taking the reflection-type metasurface as an example, and assume that the incident EM wave impinging on a unit cell is $E_{n,m}$ and the programmable reflection coefficient (i.e., a parameter that describes the complex fraction of the EM wave reflected by an impedance discontinuity in the transmission medium) of this unit cell is $A_{n,m}\cdot{e^{j\varphi_{n,m}}}$, with $A_{n,m}$ and $\varphi_{n,m}$ representing the controllable amplitude and phase shift, respectively. Then, the reflected EM wave from this unit cell can be written as

\begin{equation}\label{s1}
\tilde{E}_{n,m}=A_{n,m}\cdot{e^{j\varphi_{n,m}}}\cdot{E_{n,m}}.
\end{equation}

The total reflected EM wave $\tilde{E}$ observed at a certain location $\bf p$ in front of the metasurface is the superposition of those reflected by all the unit cells, which depends on the wireless channel $h_{n,m}(\bf{p})$ between each unit cell and the observation point, i.e.,

\begin{equation}\label{s2}
\tilde{E}({\bf p})=\sum\limits_{n = 1}^N {\sum\limits_{m = 1}^M h_{n,m}({\bf p}){\cdot}\tilde{E}_{n,m}}=\sum\limits_{n = 1}^N {\sum\limits_{m = 1}^M h_{n,m}({\bf p}){\cdot}A_{n,m}\cdot{e^{j\varphi_{n,m}}}\cdot{E_{n,m}}}.
\end{equation}

The above equation illustrates the fundamental working principle of the reflection-type metasurface, with similar principle holds for its transmission-type counterpart. The only difference is that the transmission-type one metasurface manipulates the transmitted EM wave instead of the reflected EM wave.

\section{Metasurface-based wireless transceiver}\label{Metatransceiver}
In this section, we will propose two novel tranceiver architectures enabled by programmable metasurfaces, namely metasurface-based RF chain-free transmitter and space-down-conversion receiver. Such new architectures have great potential to reduce the hardware cost and complexity of wireless transceivers in the future UM-MIMO and THz wireless communications.

\subsection{RF Chain-Free Transmitter}\label{Metatransmitter}
Wireless transmitters play a vital role in modern wireless communication systems, which have achieved great advancement in the past few decades. However, there are quite few fundamental innovations in the design of transmitter architectures despite of the rapid development of the electronic technologies. Most of today's high-performance transmitters still rely on the conventional architecture shown in Fig. 2(a), where each RF chain needs one power amplifier (PA), two mixers and several filters. This leads to extremely high hardware cost and power consumption when applied in UM-MIMO. There have been some prior research efforts to address the above issues. For example, the direct antenna modulation (DAM) technique has been proposed to directly generate modulated RF signals using time-varying antennas, which greatly simplify the hardware architecture. However, such an architecture only supports several inefficient basic modulation schemes such as on-off keying (OOK) \cite{DMA1}\cite{DMA2} and frequency shift keying (FSK) \cite{DMA3}. A similar technique has been proposed in \cite{DPSM} to utilize the direct phase shifter modulation (DPSM) method to achieve phase modulation through compact architecture, but it suffers from low transmission rate due to the slow update rate of the phase shifters.
\begin{figure}[H]
	\centering
	\includegraphics[height=2.6in]{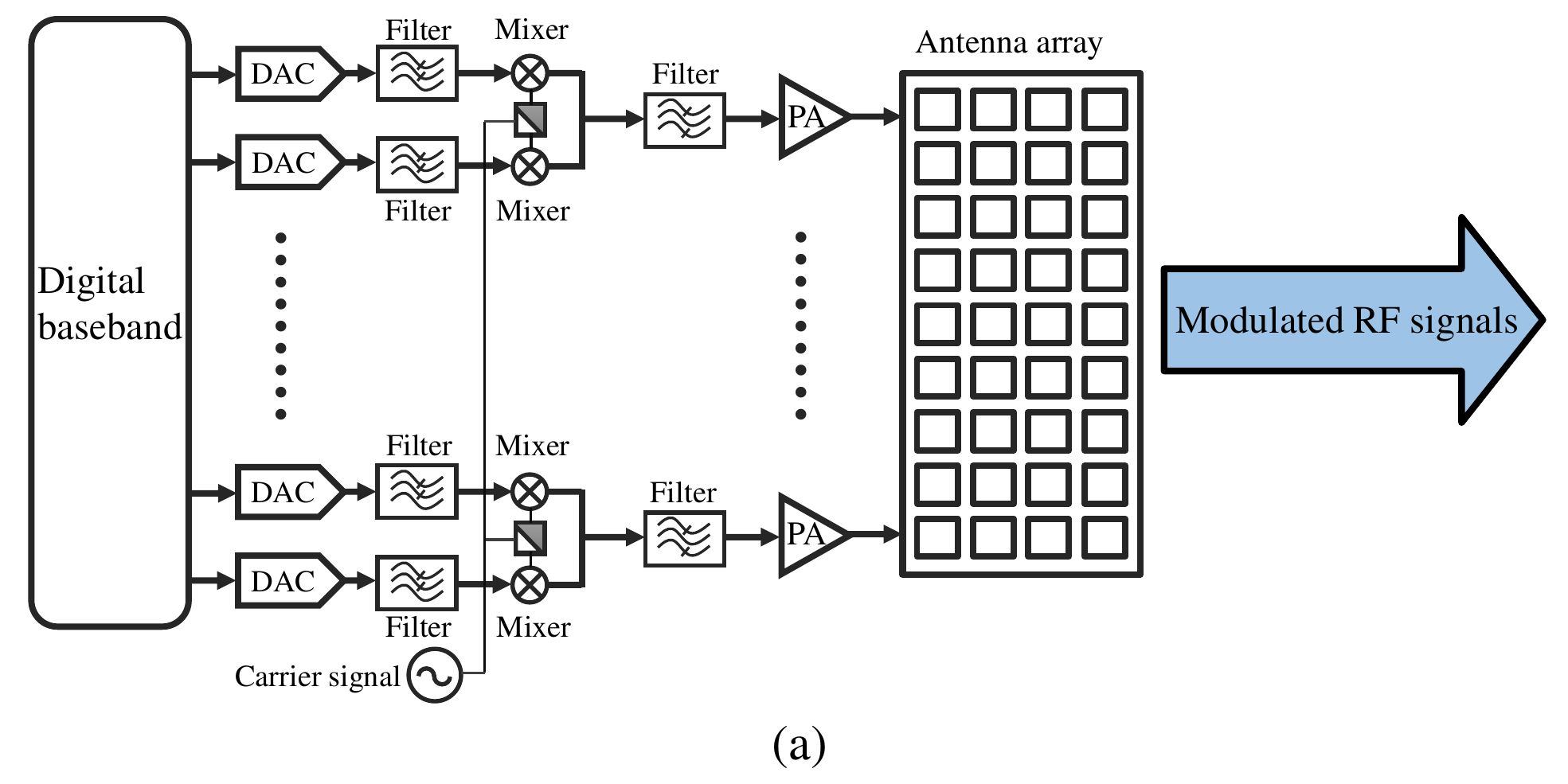}
    \includegraphics[height=2.4in]{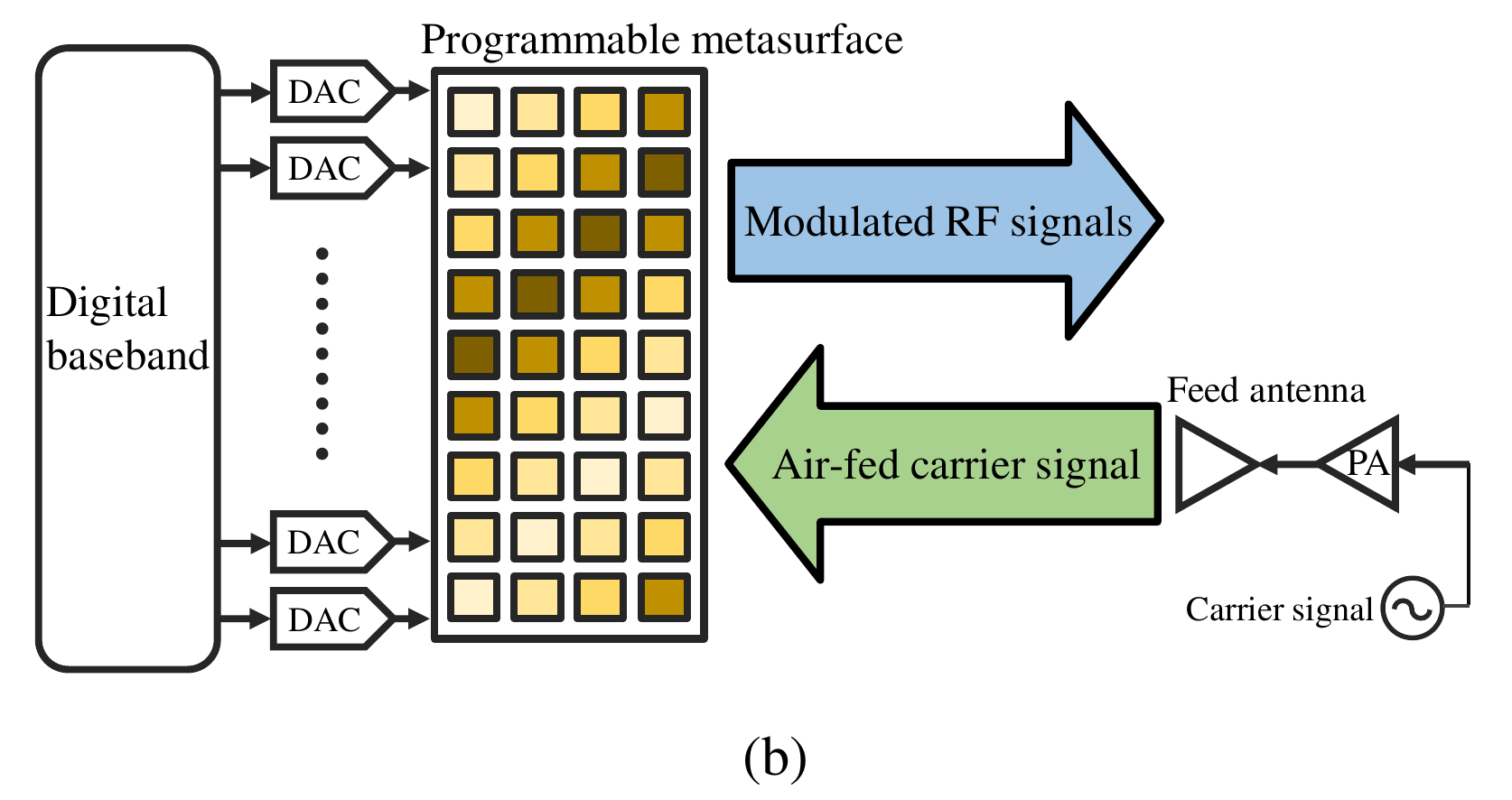}

	\caption{. Comparison of the conventional wireless transmitter (a) and the metasurface-based RF chain-free wireless transmitter (b).}
	\label{conmetatransmitter}
\end{figure}

Fig. 2(b) shows our proposed RF chain-free transmitter architecture, which aims to fundamentally reduce the hardware complexity targeting for UM-MIMO and THz communications. With this architecture, the single-tone carrier signal is fed to the radiating elements (unit cells) through the air by a feed antenna. Then the digital baseband is directly mapped to the control signals to regulate the reflection coefficients of the unit cells of the programmable metasurface, thereby achieving modulation of the reflected EM wave. For instance, phase shift keying (PSK) modulation can be achieved by applying different control signals to the programmable metasurface for different phase manipulation of the reflected RF signals \cite{MetaCom}. In principle, since the amplitude and phase response of each unit cell can be controlled independently by a dedicated digital-to-analog converter (DAC), the proposed transmitter can generate multi-channel RF signals simultaneously, which consequently enables advanced signal processing methods such as space-time modulation and beam steering for MIMO and the future UM-MIMO technologies. Meanwhile, such an RF chain-free paradigm requires only one narrow band PA to manage the transmit power of the air-fed carrier signal without the need for mixers and filters, regardless of how many channels are used \cite{MetaEL}. This thus greatly reduces the hardware complexity and implementation cost as compared with the conventional architecture in Fig. 2(a). In addition, the PA in the proposed architecture only needs to amplify the single-tone carrier signal instead of the modulated wideband signal, rendering it a promising technique to circumvent the nonlinearity issue of PAs. Furthermore, the passive feature of the programmable metasurface enables high energy efficiency and novel wearable applications \cite{MetaAI}, and its thin surface physical structure avails heat dissipation. In general, the proposed RF chain-free transmitter can effectively cut down the hardware cost, reduce the energy consumption and ease the integration procedure. While the architecture shown in Fig. 2(b) is based on the reflection-type metasurface, a similar architecture holds for the transmission-type metasurface.

\subsection{Space-Down-Conversion Receiver}\label{Metareceiver}
Conventionally, antennas are the most front-end component of the receiver to sense the weak EM waves in the space and pass the sensed signals to the subsequent modules in the receiver, including low-noise amplifiers (LNA), filters, mixers, analog-to-digital converters (ADC) and baseband signal processing modules. However, programmable metasurfaces bring new possibilities for a paradigm shift on how antennas can be used at the receiver. Due to the programmable phase gradient distribution on the surface, one possibility is to utilize the programmable metasurface as lens antenna array with reconfigurable focal point \cite{Lens}. On the other hand, if we change the phase of the transmission coefficient (taking the transmission-type metasurface as an example) of all the unit cells linearly with time and keep the amplitude constant, we can achieve the frequency down-conversion of the EM waves in the space to reduce the pressure of receiver hardware design. This is demonstrated by the following formula,

\begin{equation}\label{s3}
\hat{E}={e^{j(-2\pi/T_{meta})t}{\cdot}E},
\end{equation}
where $E$ and $\hat{E}$ represent the incident and transmitted EM waves of the programmable metasurface, respectively, $T_{meta}$ represents the period required for a linear phase change of $2\pi$, i.e., $1/T_{meta}$ characterizes the changing rate of the phase that varies linearly with time. After the EM signal passes through the metasurface, its center frequency shifts by $1/T_{meta}$, which is defined as \emph{space-down-conversion} here because the down-conversion is achieved in the space. This process works like the down-converting mixers in the conventional receiver, but it is more efficient and less complex for ultra-massive channels since only one passive metasurface is required.

\begin{figure}[H]
	\centering
	\includegraphics[width=7in]{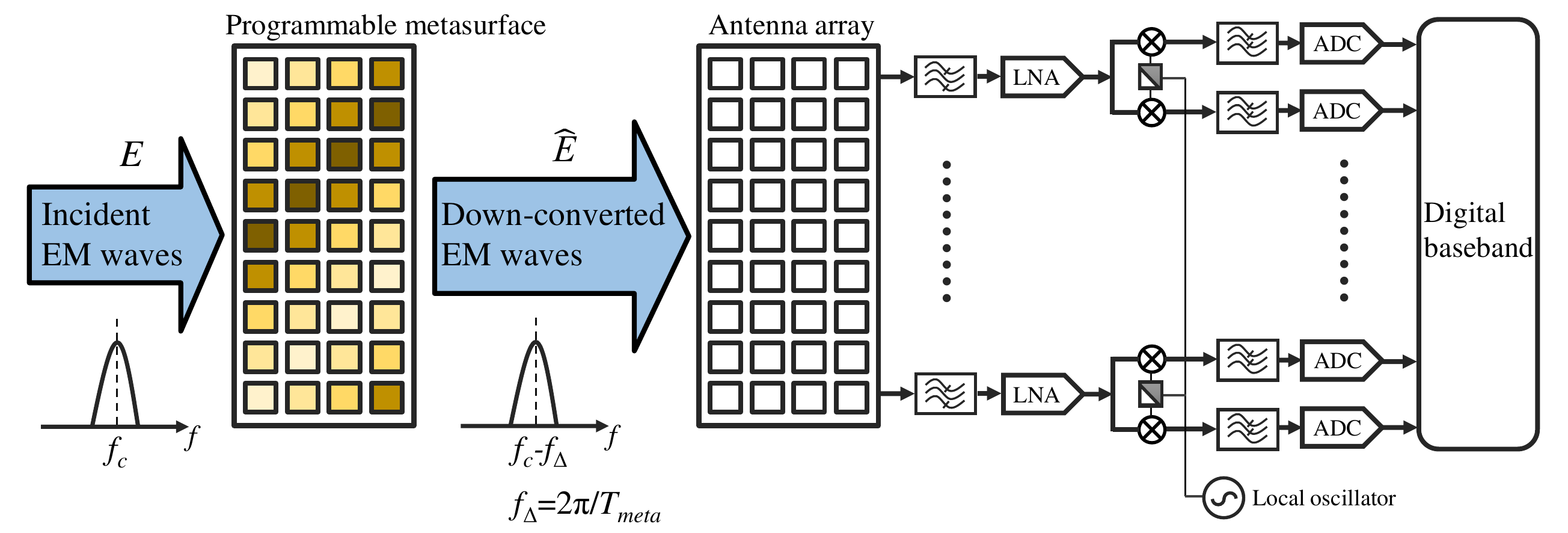}

	\caption{. Metasurface-based space-down-conversion receiver (transmission-type).}
	\label{metarx}
\end{figure}
This new paradigm is illustrated in Fig. \ref{metarx}. The high-frequency incident EM waves are converted to low-frequency EM waves after passing through the programmable metasurface, which functions as a frequency down converter in the space. Then, the low-frequency (after down-conversion) EM waves are captured by the antenna array behind the programmable metasurface, thus rendering the RF and analog circuit modules of the receiver, such as filters, LNAs, mixers and local oscillators, only need to operate at a relative low working frequency like sub-6GHz, leading to lower hardware cost. In addition, the extensively studied hybrid receiver architecture can also be combined with the proposed paradigm, to further decrease the hardware cost and complexity of the receiver with ultra-massive RF chains or high working frequency band in the future.

\subsection{Integrated Transceiver Design}\label{Integration}
Modern digital wireless communication systems work in duplex mode. For example, the BS should be capable of receiving the users' request and feedback information, and then transmit the data to the users. Therefore, it is of significant importance to consider the integrated design of the above-mentioned RF chain-free transmitter and space-down-conversion receiver. In the integrated metasurface-based transceiver design, a Tx/Rx control signal can be used to switch between transmitter and receiver modes. In the transmitter mode, the single-tone carrier signal passes through one power amplifier and connects to the antenna array (feed antenna array), and then illuminates the programmable metasurface. Meanwhile, the transmission/reflection coefficients of the unit cells on the metasurface are manipulated by the control signals, thereby realizing multi-channel transmission. In the receiver mode, the programmable metasurface is controlled by the down-conversion function signal and acts as a down-converter. At the same time, the antenna array is switched to the receiving chains, thus receiving the down-converted RF signals and then obtaining the baseband signals for processing.

\section{Test-bed setup and experimental results}\label{Testsetup}
To validate the feasibility of the metasurface-based RF chain-free transmitter and the space-down-conversion receiver described in section \ref{Metatransceiver}, we conducted the proof-of-concept experiments over the air by using the programmable metasurface, several commercial off-the-shelf modules and software defined radio (SDR) platforms. While our experiments were conducted at microwave frequencies, it can be further extended to the higher frequency band like THz in combination with the advanced manufacturing and design technologies for programmable metasurface \cite{MetaTHz}.

\begin{figure}[H]
	\centering
	\includegraphics[width=6.7in]{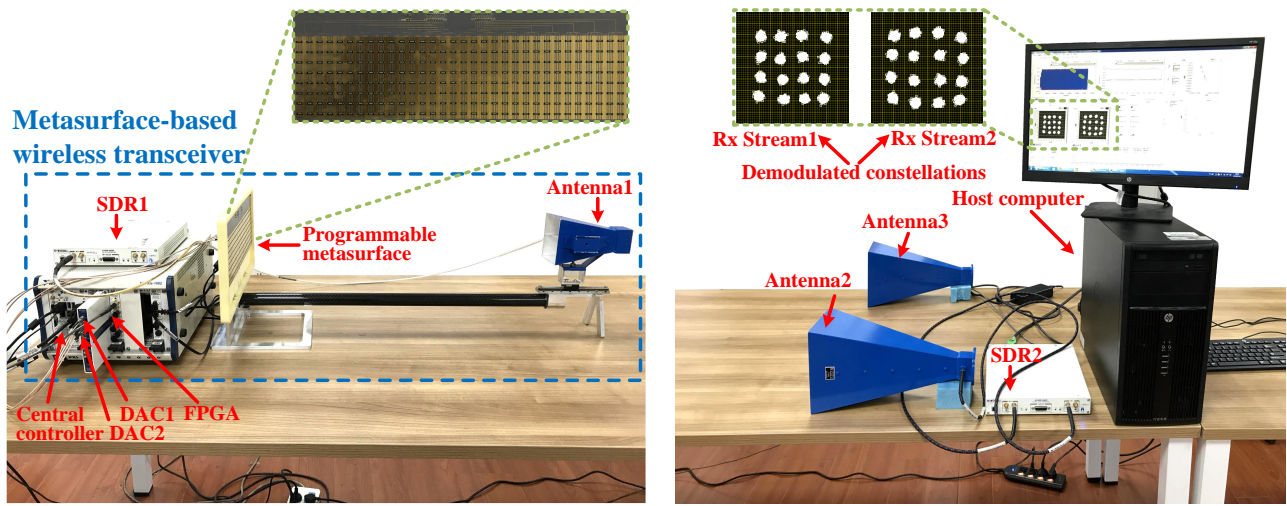}

	\caption{. The test-bed setup of the metasurface-based transceiver.}
	\label{testbed}
\end{figure}

The metasurface-based wireless transceiver in the test-bed system is shown in the left part of Fig. \ref{testbed}. It consists of the programmable metasurface, DAC modules (DAC1 and DAC2), FPGA module, central controller, horn antenna (antenna1) and SDR (SDR1). In the RF chain-free transmitter experiment, 2x2 MIMO-16QAM transmission with 20 Mbps data rate is achieved, as shown in Fig. \ref{testbed}. Two pilot and source bit streams are generated by the central controller, which are sent to the FPGA module. The FPGA module maps the bit streams into two digital baseband symbol sequences as the input of DAC1 and DAC2, respectively. The DAC1 and DAC2 convert the two digital baseband sequences into two control signal sequences for the programmable metasurface. One control signal sequence is used to control the reflection coefficients of the left half of the programmable metasurface and the other is used to control those of the right half, thus realizing the 2x2 MIMO modulation of the reflected EM waves, together with the single-tone incident carrier wave generated by the SDR1. It's important to note that in practical implementations, the incident carrier signal can be generated by using a low-cost single-tone RF signal source, rather than by the expensive SDR platform. Besides, the programmable metasurface used here has 256 unit cells. If each unit cell is controlled by a dedicated DAC, it can achieve simultaneous transmission of different signals over 256 unit cells, i.e., achieving the scale of UM-MIMO. The use of only two DACs to control the metasurface here is mainly limited by the experimental hardware conditions, but this is already sufficient to demonstrate the huge potential of the RF chain-free transmitter for UM-MIMO transmission. A conventional two-channel receiver is used to detect the signals transmitted by the programmable metasurface, which is composed of the receiving antennas (antenna2 and antenna3), the SDR platform (SDR2) and the host computer, as shown in the right part of Fig. \ref{testbed}. SDR2 passes the received baseband signals to the host computer, which then implements MIMO detection and displays the demodulated 16QAM constellation diagrams of stream1 and stream2.

\begin{figure}[H]
	\centering
	\includegraphics[width=6in]{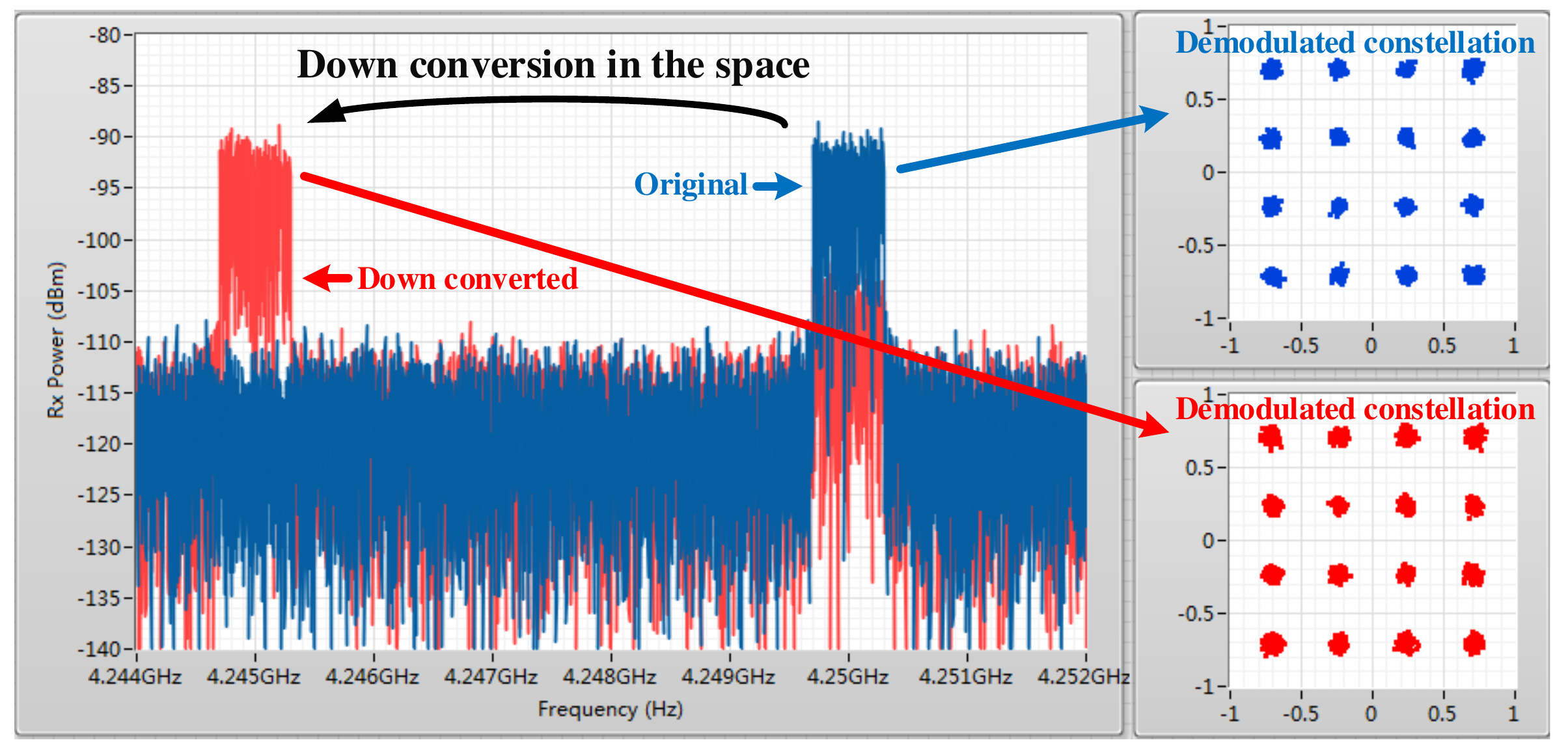}

	\caption{. Space-down-conversion receiver experimental result.}
	\label{testspacedownconversion}
\end{figure}

Table \ref{compare} compares the metasurface-based RF chain-free transmitter with the existing techniques in the literatures based on direct antenna modulation and direct phase shifter modulation. The metasurface-based technique has the advantages of supporting high-order modulations, MIMO transmission and high data rate, which are all important performance metrics of transmitters. These preliminary experimental results verify the feasibility of the RF chain-free transmitter and reveal its great potential in the future UM-MIMO and THz wireless communication systems.
\vspace{-0.5cm}
\begin{table}[H]
\centering
\footnotesize
\caption{Comparison of different methods to achieve simplified transmitter architecture.}\label{compare}
\begin{tabular}{l|l|l|l|l|l}
\hline
\textbf{Work} & \textbf{Method}  & \textbf{Modulation Scheme} & \textbf{Transmission Scheme} & \textbf{Data Rate} & \textbf{Carrier Frequency}\\
\hline
~\cite{DMA1}    & DAM & OOK & SISO & 1 kbps & 10 GHz\\
\hline
~\cite{DMA2}    & DAM & OOK & SISO & 100 Mbps & 2.4 GHz\\
\hline
~\cite{DMA3}    & DAM & BFSK & SISO & 12 Mbps & 42/57 MHz\\
\hline
~\cite{DPSM}    & DPSM & QPSK & SISO & 200 kbps & 7 GHz\\
\hline
this article & Metasurface-based & 16QAM & 2x2 MIMO & 20 Mbps & 4.25 GHz\\
\hline
\end{tabular}
\end{table}
 In the space-down-conversion receiver experiment, down conversion in the space is realized as shown in Fig. \ref{testspacedownconversion}, where the experiment is conducted in the same test-bed system presented in Fig. \ref{testbed}. SDR2 generates the 16QAM modulation RF signal at 4.25 GHz center frequency and radiates it through antenna2. If the metasurface does not perform the space-down-conversion function, then DAC1 and DAC2 provide the same fixed control signal to all the unit cells of the programmable metasurface, that is, the metasurface acts as a reflector in this case, and SDR1 receives the signal of 4.25 GHz center frequency over the air through antenna1 and demodulates it. The blue spectrum and constellation diagram shown in Fig. \ref{testspacedownconversion} are the corresponding received spectrum and demodulated constellation. When the programmable metasurface acts as a space-down-converter, the DAC1 and DAC2 provide a carefully designed continuous control signal sequence to all the unit cells of the programmable metasurface, which makes their reflection phase responses change linearly with time, as described in section \ref{Metatransceiver}. In our preliminary experiment, 5 MHz space-down-conversion is achieved, as shown in Fig. \ref{testspacedownconversion}. The original blue RF signal is shifted to the down-converted red RF signal after passing through the metasurface, which makes the SDR1 in the metasurface-based transceiver only need to deal with the 4.245 GHz signal instead of the 4.25 GHz signal. The demodulated red constellation diagram after space-down-conversion is as good as the original blue one in Fig. \ref{testspacedownconversion}. This result verifies the feasibility of the metasurface-based space-down-conversion receiver paradigm. At the moment, only 5 MHz frequency down conversion is realized, as the sampling rate of the DAC modules used in the experiment is only 100 MHz, and the control sequence here for completing the $2\pi$ phase linear change with time is generated by 20 sampling points, so $1/T_{meta}$ mentioned in section \ref{Metatransceiver} can only reach 5 MHz. If dedicated circuit for ultra-fast control sequence generation is designed and the phase regulation speed of the programmable metasurface is fast enough, GHz and even THz level space-down-conversion can be potentially implemented in the future, which will greatly reduce the hardware complexity and the cost of the UM-MIMO or THz receiver in 6G era.

\section{Challenges and future research directions}\label{Challenges}

In this article, we have introduced two new wireless communication paradigms utilizing programmable metasurfaces. Their main advantages are discussed, together with some preliminary experimental results. In the following, we outline some main challenges and future research directions for metasurface-based wireless transceiver.

\subsection{Theoretical Modeling}

As the novel architecture of metasurface-based RF chain-free transmitter and space-down-conversion receiver is significantly different from the conventional ones, it is necessary to develop analytical signal models and electromagnetic field models to pave the way for further theoretical studies. The non-ideal hardware characteristics of the programmable metasurface, such as the nonlinearity of phase response and the effect of charge/discharge of the tunable components in the unit cells, should be taken into account for the theoretical modeling. Moreover, the space-down-conversion paradigm forms a new mode of cascading high-frequency and low-frequency channels together through the metasurface. How to reasonably model these new modes and their characteristics is the key to study the channel capacity, spectrum efficiency and transmission schemes of the metasurface-based transceivers.

\subsection{Transceiver Scheme Design}

The current work of metasurface-based transceiver only considers single carrier modulation, whereas realizing OFDM with programmable metasurface is more challenging. It is necessary to study how to combine metasurface with OFDM-MIMO technology. Exploring new mapping methods from OFDM-MIMO baseband to reflection/transmission coefficients of the programmable metasurface could make it possible to tremendously improve the transmission rate and spectrum utilization. In addition, the combination of metasurface and some cutting-edge transmission technologies is also promising, such as orbital angular momentum (OAM). Moreover, the working paradigm of the metasurface in the receiver needs further exploration. How to realize direct demodulation based on the programmable metasurface is a big challenge, and once realized, it will subvert the working principle of the conventional receivers and bring extraordinary technical breakthroughs.

\subsection{Practical Measurement}

Since the research of applying the metasurface in the wireless transceiver is still in its early stage, practical measurement work is quite necessary and important, which may reveal the fundamental differences between the new metasurface-based and the conventional transceiver architectures through measurement data, and provide a powerful foundation for theoretical modeling. Some possible measurement directions include metasurface array gain, beam steering, conversion efficiency, path loss measurements, and more.

\subsection{Prototyping Work}

The practical performance of the metasurface-based wireless transceiver depends on the design level and manufacturing technique of the programmable metasurfaces, such as their regulation speed and manipulation accuracy on the EM waves. With the development of the metasurface technology, the performance of metasurface-based wireless transceivers will continue to improve in the future. Prototyping works with the most advanced metasurfaces would better showcase the progress in this filed and provide design reference to the researchers.

\section{Conclusion}\label{Conclusion}

In this article, we have proposed two paradigms to utilize the programmable metasurfaces as RF chain-free transmitter and space-down-conversion receiver, which both have great potential to realize cost-effective and energy-efficient wireless communication networks in the future. The basic principles, design architectures and promising advantages of these new paradigms are presented. It is demonstrated that programmable metasurfaces may bring a paradigm shift on the wireless transceiver design, due to their superior capability of manipulating the EM waves. The future research directions are also discussed, which include theoretical modeling, performance analysis, transceiver scheme optimization, system prototyping and over the air measurements.

\end{document}